\documentclass[conference]{IEEEtran}

\usepackage{amsmath}
\usepackage{amssymb}
\usepackage[dvips]{graphicx}
\usepackage{epsfig}

\newcommand{\SNR}{{\text{SNR}}}

\newtheorem{lemma:bitenergylow}{Lemma}
\newtheorem{lemma:bitenergyhigh}[lemma:bitenergylow]{Lemma}

\newtheorem{prop:asympcap}{Theorem}
\newtheorem{prop:flashminbitenergy}[prop:asympcap]{Theorem}
\newtheorem{prop:flashbitenergy}[prop:asympcap]{Theorem}
\newtheorem{prop:pasympcap}[prop:asympcap]{Theorem}

\begin{document}


\title{Low-SNR Analysis  of  Interference Channels under Secrecy Constraints}



%
\author{\authorblockN{Junwei Zhang and Mustafa Cenk Gursoy}
\authorblockA{Department of Electrical Engineering\\
University of Nebraska-Lincoln, Lincoln, NE 68588\\ Email:
junwei.zhang@huskers.unl.edu, gursoy@engr.unl.edu}}


\maketitle
\begin{abstract}
In this paper, we study the secrecy rates over weak Gaussian
interference channels for different transmission schemes. We focus on the low-SNR regime and obtain the minimum bit energy $\frac{E_b}{N_0}_{\min}$ values, and the wideband slope regions for both TDMA and multiplexed transmission
schemes. We show that secrecy constraints introduce
a penalty in both the minimum bit energy and the slope regions. Additionally, we identify
under what conditions TDMA or multiplexed transmission is
optimal. Finally, we show that TDMA is more likely to be optimal in the presence of secrecy constraints.

\end{abstract}

\section{Introduction}
The open nature of wireless communications allows for the signals to
be received by all users within the communication range. Thus,
wireless communication is vulnerable to eavesdropping. This problem
was first studied in \cite{wyner} where Wyner proposed a  wiretap channel model.
In this model, a single source-destination communication link is
eavesdropped by a wiretapper. The secrecy level is measured by the
equivocation rate. Wyner showed that secure communication is possible
without sharing a secret key if the eavesdropper's channel is
a degraded version of the main chain. Later, Wyner's result was extended to
the Gaussian channel in \cite{cheong} and recently to fading
channels in \cite{Gopala}. In addition to the single antenna case,
secrecy in multi-antenna models is addressed
in \cite{shafiee} -- \cite{Oggier}. Multiple access channels with confidential
messages were considered in \cite{liang}.  Liu  \emph{et al.} \cite{Liu} presented
inner and outer bounds on secrecy capacity regions for broadcast and
interference channels. They also described several transmission
schemes for Gaussian interference channels and derived their
achievable rate regions while ensuring mutual information-theoretic
secrecy. Recently, Bloch \emph{et al.} in \cite{bloch} discussed the theoretical aspects
and practical schemes for wireless information-theoretic security.

Another important concern in wireless communications is the
efficient use of limited energy resources. Hence, the energy
required to reliably send one bit is a metric that can be adopted to
measure the performance. Generally, energy-per-bit requirement is
minimized, and hence the energy efficiency is maximized, if the
system operates in the low-SNR regime. In \cite{verdu}, Verdu has
analyzed the tradeoff between the spectral efficiency and bit energy
in the low-SNR regime for a general class of channels. As argued
in \cite{verdu}, two key performance measures in the low-power regime are the
minimum energy per bit $\frac{E_b}{N_0}_{\min}$ required for reliable
communication and the slope of the spectral efficiency versus
$\frac{E_b}{N_0}$ curve at $\frac{E_b}{N_0}_{\min}$. Caire \emph{et al.} in
\cite{Caire} employed these two measures to study the multiple access,
broadcast, and interference channels in the low-power regime. By
comparing the performance of TDMA and superposition schemes, they
concluded that the growth of TDMA-achievable rates with energy per bit is
suboptimal except in some special cases.

In this paper, we study secure transmission over  Gaussian weak
interference channels in the low-power regime. The organization of the rest of the paper is as
follows. In Section \ref{model}, we describe the channel model and
obtain the secrecy achievable rate regions for TDMA, multiplexed
transmission schemes and artificial noise schemes, and compare
their performances in terms of the achievable rates. In Section \ref{energy}, we compute the
minimum energy per bit and slope at $\frac{E_b}{N_0}_{\min}$ for
TDMA and multiplexed transmission schemes. In Section \ref{discuss}, we
use results in Section \ref{energy} to evaluate how secrecy
constraints affect the performance in the low-power regime and identify
optimal transmission schemes. Finally, we provide
conclusions in Section \ref{conclu}.

\section{Gaussian Interference Channels with Confidential
Messages}\label{model}

We consider secure communication over a two-transmitter, two-receiver Gaussian
interference channel. The input-output relations for this channel model are given by
\begin{align}
y_1&=c_{11}x_1+c_{12}x_2+n_1, \text{ and }\\
y_2&=c_{21}x_1+c_{22}x_2+n_2
\end{align}
where $x_1$ and $x_2$ are the channel inputs of the transmitters, the coefficients $\{c_{ij}\}$ denote the channel gains and are deterministic
scalars, and $n_1$ and $n_2$ are independent, circularly symmetric, complex Gaussian random variables with zero mean and
common variance $\sigma^2$. It is assumed that the transmitters are subject to the following average power constraint:
\begin{align}
E[|x_i |^2] \leqslant P_i = \SNR_i \,\sigma^2 ,~~~~i=1,2.
\end{align}
We focus on the weak interference channel i.e., we assume that
$\frac{|c_{12}|^2}{|c_{11}|^2}<1$ and $\frac{|c_{21}|^2}{|c_{22}|^2}<1$.
Over this channel, transmitter $i$ for $i=1,2$ intends to send an confidential message by transmitting $x_i$ to the desired
 receiver $i$, which receives $y_i$, while ensuring that the other receiver does not
 obtain any information by listening the transmission. Following \cite{Liu},
 we next consider three transmission schemes and their corresponding achievable secrecy rate regions.

\subsection{Time Division Multiple Access}

In TDMA, the transmission period is divided into two
nonoverlapping time slots. Transmitters 1 and 2 transmit using $\alpha$ and $1-\alpha$ fractions of time, respectively.
We note that
under this assumption, the channel in each time slot reduces to a
Gaussian wiretap channel \cite{cheong}, and the following rate region
can be achieved with perfect secrecy \cite{Liu}:
\begin{align}\label{tdma}
&R_1\geqslant 0\nonumber \\
&R_2\geqslant 0 \nonumber \\
 &R_1\leqslant \alpha\left[\log\left(1+\frac{|c_{11}|^2
\SNR_1}{\alpha}\right)-\log\left(1+\frac{|c_{21}|^2 \SNR_1}{\alpha}\right)\right]\nonumber\\
&R_2\leqslant (1-\alpha)\left[\log\left(1+\frac{|c_{22}|^2
\SNR_2}{1-\alpha}\right)-\log\left(1+\frac{|c_{12}|^2 \SNR_2}{1-\alpha}\right)\right]
\end{align}
over all possible transmitting signal-to-noise-ratio pairs $\SNR_1\in
[0,P_1/\sigma^2], \SNR_2 \in [0,P_2/\sigma^2]$ and time allocation
parameter $\alpha$.
\subsection{Multiplexed Transmission} In the multiplexed
transmission scheme, transmitters are allowed to share the same
degrees of freedom. By the constraint of information-theoretic
security, no partial decoding of the other transmitter's message is
allowed at a receiver. Hence, the interference results in an
increase of the noise floor. Thus, the following rate region can be
achieved with perfect secrecy \cite{Liu}:
\begin{align}\label{multi}
&R_1\geqslant 0 \nonumber\\
&R_2\geqslant 0 \nonumber \\
&R_1\leqslant \log\left(1+\frac{|c_{11}|^2 \SNR_1}{1+|c_{12}|^2
\SNR_2}\right)-\log\left(1+|c_{21}|^2 \SNR_1\right) \nonumber \\
&R_2\leqslant \log\left(1+\frac{|c_{22}|^2 \SNR_2}{1+|c_{21}|^2
\SNR_1}\right)-\log(1+|c_{12}|^2 \SNR_2)
\end{align}
over all possible transmitting signal-to-noise-ratio pairs $\SNR_1\in
[0,P_1/\sigma^2], \SNR_2 \in [0,P_2/\sigma^2]$.

\subsection{Artificial Noise}
This scheme allows one of the transmitters (e.g transmitter 2) to generate
artificial noise. This scheme will split the power of transmitter 2 into
two parts: $\lambda P_2$ for generating artificial noise and the
remaining $(1-\lambda)P_2$ for encoding the confidential message. As
detailed in \cite{Liu}, the achievable rate region is
\begin{align}\label{arti}
&R_1\geqslant 0 \nonumber\\
&R_2\geqslant 0 \nonumber \\
&R_1\leqslant \log\left(1+\frac{|c_{11}|^2 \SNR_1}{1+|c_{12}|^2
\SNR_2}\right)\!\!-\!\!\log\left(1+\frac{|c_{21}|^2 \SNR_1}{1+|c_{22}|^2 \lambda \SNR_2}\right) \nonumber \\
&R_2\leqslant \log\left(1+\frac{|c_{22}|^2 (1-\lambda)\SNR_2}{1+|c_{21}|^2
\SNR_1+|c_{22}|^2\lambda \SNR_2}\right)\nonumber\\
& \hspace{1cm}-\log\left(1+\frac{|c_{12}|^2 (1-\lambda)\SNR_2}{1+|c_{12}|^2 \lambda
\SNR_2}\right)
\end{align}
over all possible transmitting signal-to-noise-ratio pairs $\SNR_1\in
[0,P_1/\sigma^2], \SNR_2 \in [0,P_2/\sigma^2] $ and power splitting
parameter $\lambda$. We can further enlarge the rate region by
reversing the roles of transmitters 1 and 2.

When the transmitting power is moderate, neither too high nor too
small, as demonstrated in \cite{Liu}, transmission strategy with artificial noise provides the
largest achievable rate region while TDMA gives the smallest rate region.

On the other hand, when we consider the two extreme cases of high- and low-SNR regimes, the picture changes. In the high-SNR regime,
when we let $\SNR_1 \to \infty, \SNR_2 \to \infty $ and
$\lim\frac{\SNR_1}{\SNR_2}=q$ in
(\ref{tdma}), (\ref{multi}), and (\ref{arti}), we can see that multiplexed
transmission can not achieve any positive secrecy rate, while TDMA rates are
bounded by $R_1<\alpha\log(\frac{|c_{11}|^2}{|c_{21}|^2})$, and $R_2<
(1-\alpha)\log(\frac{|c_{22}|^2}{|c_{12}|^2})$. For the strategy with the artificial noise,
rate $R_1$ is bounded by
$R_1<\log(\frac{1+\frac{|c_{11}|^2q}{|c_{12}|^2}}{1+\frac{|c_{21}|^2q}{|c_{22}|^2\lambda}})$,
but we can not achieve any secrecy rate for $R_2$. Thus, TDMA is the best choice when we want both users to have secure communication in the high-SNR regime.

\begin{figure}
\begin{center}
\includegraphics[width = 0.45\textwidth]{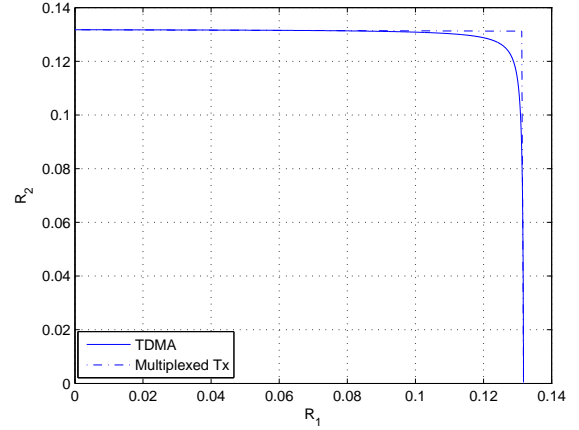}
\caption{Gaussian Interference Channel secrecy rate achievable
Region $P_1=P_2=0.1,c_{11}=c_{22}=1,c_{12}=c_{21}=0.2$}
\label{fig:region01}
\end{center}
\end{figure}

In the low-SNR regime (as $\SNR$ approaches zero), TDMA and multiplexed
transmission achievable regions become identical. They
converge to the following rectangular rate region, as illustrated in
Fig.\ref{fig:region01}:
\begin{align}\label{limit}
&R_1\geqslant 0\nonumber \\
&R_2\geqslant 0 \nonumber \\
&R_1\leqslant |c_{11}|^2 \SNR_1-|c_{21}|^2 \SNR_1+o(\SNR_1) \nonumber \\
&R_2\leqslant |c_{22}|^2 \SNR_2-|c_{12}|^2 \SNR_2+o(\SNR_2)
\end{align}
Thus, these schemes have similar performances at vanishing SNR levels in terms of the asymptotic rates. However, a finer analysis in the next section will provide more insight.  We note that in the case of transmission with artificial noise, we have
$R_1\leqslant |c_{11}|^2 \SNR_1-|c_{21}|^2 \SNR_1 + o(\SNR_1)$ and
$R_2\leqslant (1-\lambda)(|c_{22}|^2 \SNR_2-|c_{12}|^2 \SNR_2) + o(\SNR_2)$ which
is strictly smaller than that in (\ref{limit}). This lets us to conclude that introducing artificial noise is not preferable in the low-SNR regime.

\section{energy efficiency in the low-snr regime}\label{energy}
The tradeoff of spectral efficiency versus energy per information
bit is the key measure of performance in the low-SNR regime.
The two major analysis tools in this regime are the
minimum value of the energy per bit $\frac{E_b}{N_0}_{\min}$, and the
slope $S$ of the spectral efficiency versus $\frac{E_b}{N_0}$ curve
at $\frac{E_b}{N_0}_{\min}$ \cite{verdu}. These can be obtained from
\begin{align}
\frac{E_b}{N_0}_{\min}=\frac{\log_e 2}{\dot{C}(0)} \label{ebno} \\ \intertext{and}
S=  \frac{2[\dot{C}(0)^2]}{-\ddot{C}(0)}\label{slope}
\end{align}
where $\dot{C}(0)$ and $\ddot{C}(0)$ denote the first and second derivatives of the channel capacity with respect to SNR at SNR = 0.

In this section, using these tools, we analyze the performance in interference channels with confidential messages, following an approach similar to that in \cite{Caire}. Note that in interference channels, we have the achievable rate pairs
$(R_1,R_2)$. As the SNRs of both users approach zero in the low-SNR regime, it can be easily seen that $R_1 \to 0$ and $R_2 \to 0$.  In this regime, we introduce the parameter $\theta$, and assume that the ratio of the rates is $R_1/R_2=\theta$ as $R_1$ and $R_2$ both vanish. In
both TDMA and multiplexed transmissions, we have
\begin{align}
\theta=\frac{R_1}{R_2}=\frac{\SNR_1(|c_{11}|^2-|c_{21}|^2)}{\SNR_2(|c_{22}|^2-|c_{12}|^2)}.
\end{align}
By fixing $\theta$, we can rewrite the achievable rate region of
multiplexed transmission in (\ref{multi}) as
\begin{align}\label{newmulti}
&R_1\geqslant 0\nonumber \\
&R_2\geqslant 0 \nonumber \\
&R_1\leqslant \log
\left(1+\frac{|c_{11}|^2
\SNR_1}{1+|c_{12}|^2\frac{(|c_{11}|^2-|c_{21}|^2)}{\theta(|c_{22}|^2-|c_{12}|^2)}
\SNR_1}\right)\nonumber\\
&-\log(1+|c_{21}|^2 \SNR_1) \nonumber \\
&R_2\leqslant \log\left(1+\frac{|c_{22}|^2
\SNR_2}{1+|c_{21}|^2\frac{\theta(|c_{22}|^2-|c_{12}|^2)}{(|c_{11}|^2-|c_{21}|^2)}
\SNR_2}\right)\nonumber\\
&-\log(1+|c_{12}|^2 \SNR_2).
\end{align}
>From (\ref{tdma}) and (\ref{newmulti}), we can see that when SNR diminishes, the bit energy $\frac{E_b}{N_0}=\frac{\SNR}{R(\SNR)}$
for both TDMA and multiplexed transmission schemes monotonically decreases.  Furthermore, it can be shown that the rates are concave functions
of $\SNR$ in the low-SNR regime. Thus, the minimum energy per bit is achieved
as $\SNR \to 0$. The following theorems provide the minimum energy per
bit and the slope at the minimum energy per bit.

\begin{prop:asympcap} \label{prop:ebno}
For all $\theta=R_1/R_2$, the minimum bit energies in the Gaussian interference channel with confidential messages for both TDMA and
multiplexed transmissions are
\begin{align}
&\frac{E_1}{N_0}_{\min}=\frac{\log_e 2}{|c_{11}|^2-|c_{21}|^2}, \label{ebno1} \\
&\frac{E_2}{N_0}_{\min}=\frac{\log_e
2}{|c_{22}|^2-|c_{12}|^2}.\label{ebno2}
\end{align}
\end{prop:asympcap}

\emph{Proof}: From (\ref{tdma}) and (\ref{newmulti}), we can for both cases easily compute the derivatives of the achievable rates with respect to SNR as
\begin{align}
\dot{R}_1(0)=|c_{11}|^2-|c_{21}|^2\\
\dot{R}_2(0)=|c_{22}|^2-|c_{12}|^2.
\end{align}
Using (\ref{ebno}), we get the minimum bit energy expressions. \hfill
$\square$

>From the result of Theorem \ref{prop:ebno}, we see that TDMA and multiplexed
transmission achieve the same minimum energy per bit. Next, we consider the wideband slope regions.

\begin{prop:asympcap} \label{prop:slope}
Let the rates vanish while keeping $R_1/R_2=\theta$. Then, for the Gaussian interference channel with confidential messages,  the slope region achieved by TDMA is
\begin{align}\label{slopetdma}
&0\leqslant S_1<2\nonumber \\
&0\leqslant S_2<2\nonumber \\
&\frac{S_1}{2A}+\frac{S_2}{2B}=1
\end{align}
and the slope region achieved by multiplexed transmission is
\begin{align}\label{slopemul}
&0\leqslant S_1<2\nonumber \\
&0\leqslant S_2<2\nonumber \\
\left(\frac{2A}{S_1}-1\right) \left(\frac{2B}{S_2}-1\right)
&=\frac{4|c_{11}|^2|c_{12}|^2|c_{22}|^2|c_{21}|^2}{(|c_{11}|^4-|c_{21}|^4)(|c_{22}|^4-|c_{12}|^4)}
\end{align}
where
\begin{align}
A=\frac{|c_{11}|^2-|c_{21}|^2}{|c_{11}|^2+|c_{21}|^2},\\
B=\frac{|c_{22}|^2-|c_{12}|^2}{|c_{22}|^2+|c_{12}|^2}.
\end{align}
\end{prop:asympcap}

\emph{Proof}: Note again that for both transmission schemes, we have
\begin{align}
\dot{R}_1(0)=|c_{11}|^2-|c_{21}|^2,\\
\dot{R}_2(0)=|c_{22}|^2-|c_{12}|^2.
\end{align}
In TDMA, we also have
\begin{align}
-\ddot{R}_{1}(0)=\frac{|c_{11}|^4-|c_{21}|^4}{\alpha},\\
-\ddot{R}_{2}(0)=\frac{|c_{22}|^4-|c_{12}|^4}{(1-\alpha)}.
\end{align}
Then, using (\ref{slope}), we get
 \begin{align}
 S_1&=\frac{2\alpha(|c_{11}|^2-|c_{21}|^2)}{|c_{11}|^2+|c_{21}|^2},\\
 S_2&=\frac{2(1-\alpha)(|c_{22}|^2-|c_{12}|^2)}{|c_{22}|^2+|c_{12}|^2}.
 \end{align}
Considering different values of $\alpha$ leads to the region in (\ref{slopetdma}).
Similarly, for
 multiplexed transmission, we can obtain
\begin{align}
-\ddot{R}_{1}(0)&=|c_{11}|^4-|c_{21}|^4+\frac{2|c_{11}|^2|c_{12}|^2(|c_{11}|^2-|c_{21}|^2)}{\theta(|c_{22}|^2-|c_{12}|^2)},\\
-\ddot{R}_{2}(0)&=|c_{22}|^4-|c_{12}|^4+\frac{2|c_{22}|^2|c_{21}|^2\theta(|c_{22}|^2-|c_{12}|^2)}{|c_{11}|^2-|c_{21}|^2}.
\end{align}
>From the above expression, we can easily see that
\begin{align}
S_1&=\frac{2(|c_{11}|^2-|c_{21}|^2)}{|c_{11}|^2+|c_{21}|^2+\frac{2|c_{11}|^2|c_{12}|^2}{\theta(|c_{22}|^2-|c_{12}|^2)}},\\
 S_2&=\frac{2(|c_{22}|^2-|c_{12}|^2)}{|c_{22}|^2+|c_{12}|^2+\frac{2|c_{22}|^2|c_{21}|^2\theta}{|c_{11}|^2-|c_{21}|^2}}.
\end{align}
Considering different values of $\theta$ leads to the slope region given in (\ref{slopemul}). \hfill $\square$

\section{the impact of secrecy on energy efficiency}\label{discuss}
\begin{figure}
\begin{center}
\includegraphics[width = 0.4\textwidth]{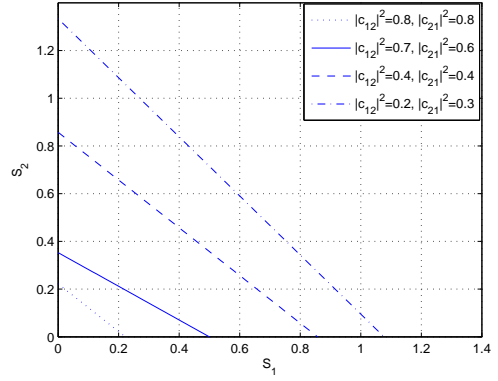}
\caption{Slope regions in the Gaussian interference channel with confidential messages for
the TDMA scheme with $|c_{11}|^2=|c_{22}|^2=1$ and various values of $|c_{12}|^2,
|c_{21}|^2$}. \label{fig:slopetd}
\end{center}
\end{figure}

\begin{figure}
\begin{center}
\includegraphics[width = 0.4\textwidth]{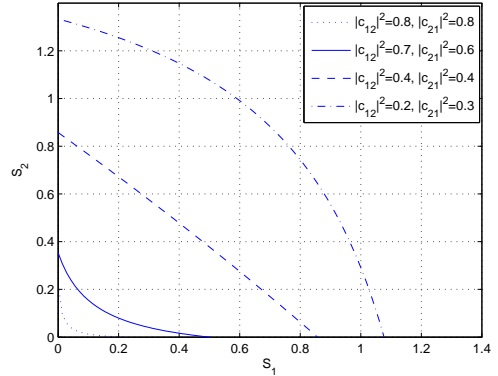}
\caption{Slope regions in the Gaussian interference channel with confidential messages for multiplexed transmission scheme with $|c_{11}|^2=|c_{22}|^2=1$ and various values of
$|c_{12}|^2, |c_{21}|^2$}. \label{fig:slopeml}
\end{center}
\end{figure}

For comparison, we provide below the minimum energy per bit and slope region
when there are no secrecy constraints \cite{Caire}. The minimum bit energies for both TDMA and multiplexed transmission are
\begin{align}
&\frac{E_1}{N_0}_{\min}=\frac{\log_e 2}{|c_{11}|^2}, \label{enbono} \\
&\frac{E_2}{N_0}_{\min}=\frac{\log_e 2}{|c_{22}|^2}.\label{ebnono1}
\end{align}
The achievable slope region for TDMA is
\begin{align}\label{slopetdmanosec}
&0\leqslant S_1<2\nonumber \\
&0\leqslant S_2<2\nonumber \\
&S_1+S_2=2,
\end{align}
while for multiplexed transmission, we have
\begin{align}\label{slopemulnosec}
&0\leqslant S_1<2\nonumber \\
&0\leqslant S_2<2\nonumber \\
&(\frac{2}{S_1}-1)(\frac{2}{S_2}-1)=4\frac{|c_{12}|^2}{|c_{22}|^2}\frac{|c_{21}|^2}{|c_{11}|^2}.
\end{align}
We can immediately note that the minimum bit energies in (\ref{enbono}) and (\ref{ebnono1}) are strictly smaller
than those given in (\ref{ebno1}) and (\ref{ebno2}). Thus, there is an energy penalty
associated with secrecy.  Moreover, comparing the slope regions in
(\ref{slopetdma}) and (\ref{slopemul}) with those in
(\ref{slopetdmanosec}) and (\ref{slopemulnosec}), and noting that
\begin{align}\label{ii}
A&<1\nonumber\\
B&<1\nonumber \\
4\frac{|c_{12}|^2}{|c_{22}|^2}\frac{|c_{21}|^2}{|c_{11}|^2}&<\frac{4|c_{11}|^2|c_{12}|^2|c_{22}|^2|c_{21}|^2}{(|c_{11}|^4-|c_{21}|^4)(|c_{22}|^4-|c_{12}|^4)},
\end{align}
we can easily verify that the slope region of Gaussian weak interference
channel is strictly larger than  the slope region of Gaussian
weak interference channel with confidential messages for both TDMA and multiplexed transmission
schemes. Thus, in addition to the increase in the minimum energy per bit, secrecy introduces a penalty in terms of the achievable wideband slope values. In Figs. \ref{fig:slopetd} and \ref{fig:slopeml}, we
plot the slope regions for TDMA and multiplexed transmissions, respectively, under secrecy constraints. We note that regions become smaller as $|c_{12}|^2$ and
$|c_{21}|^2$ increase. This is due to the fact that for fixed $|c_{11}|^2$ and
$|c_{22}|^2$, the larger values of $|c_{12}|^2$ and $|c_{21}|^2$ mean that channel of the unintended
receiver gets stronger and we have to use more energy to
achieve the same  secrecy transmission rate.

\begin{figure}
\begin{center}
\includegraphics[width = 0.4\textwidth]{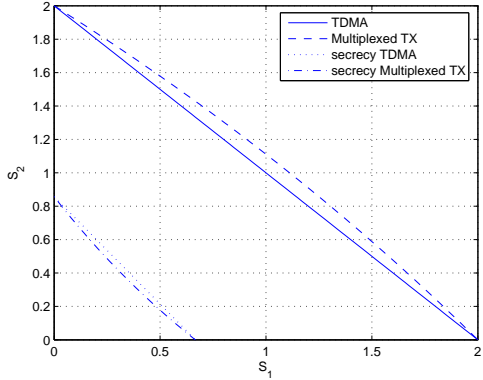}
\caption{Slope regions in the Gaussian interference channel.
$|c_{11}|^2=|c_{22}|^2=1,|c_{12}|^2=0.4, |c_{21}|^2=0.5$}
\label{fig:slope}
\end{center}
\end{figure}

\begin{figure}
\begin{center}
\includegraphics[width = 0.4\textwidth]{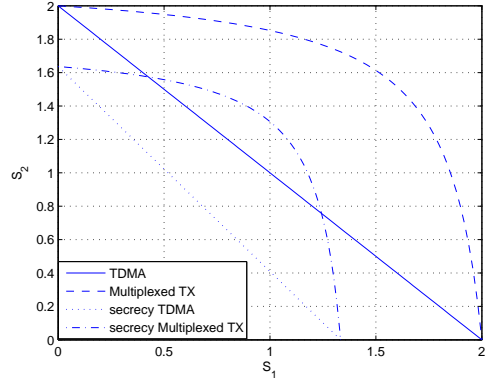}
\caption{Slope regions in the Gaussian interference channel.
$|c_{11}|^2=|c_{22}|^2=1,|c_{12}|^2=0.1, |c_{21}|^2=0.2$}
\label{fig:slope1}
\end{center}
\end{figure}

We are also interested in determining which transmission scheme performs better in the low-SNR regime. TDMA achievable rate regions converge to those of multiplexed
transmission scheme as power decreases. Furthermore, TDMA and
multiplexed transmission has the same minimum energy per bit values. Therefore, we
should consider the slope regions. From Theorem \ref{prop:slope}, we
know that when
\begin{align}
\frac{4|c_{11}|^2|c_{12}|^2|c_{22}|^2|c_{21}|^2}{(|c_{11}|^4-|c_{21}|^4)(|c_{22}|^4-|c_{12}|^4)}<1,
\end{align}
the slope region of multiplexed transmission is strictly larger
than the slope region of TDMA, thus in this case, multiplexed transmission is
preferred. On the other hand, when
\begin{align}
\frac{4|c_{11}|^2|c_{12}|^2|c_{22}|^2|c_{21}|^2}{(|c_{11}|^4-|c_{21}|^4)(|c_{22}|^4-|c_{12}|^4)}>1,
\end{align}
the slope region of TDMA is larger than the slope region of
multiplexed transmission. Hence, TDMA should be used in this scenario.
Finally, when
\begin{align}
\frac{4|c_{11}|^2|c_{12}|^2|c_{22}|^2|c_{21}|^2}{(|c_{11}|^4-|c_{21}|^4)(|c_{22}|^4-|c_{12}|^4)}=1,
\end{align}
the slope regions of TDMA and multiplexed transmission
converge to the same triangular region. In this case, TDMA should
still be preferred due to its implementational advantages.  These results show parallels to those obtained in \cite{Caire} in the absence of secrecy constraints. In \cite{Caire}, the function that is compared with one is
$4\frac{|c_{12}|^2}{|c_{22}|^2}\frac{|c_{21}|^2}{|c_{11}|^2}$. From (\ref{ii}), we see that when we vary the
channel parameters,
$\frac{4|c_{11}|^2|c_{12}|^2|c_{22}|^2|c_{21}|^2}{(|c_{11}|^4-|c_{21}|^4)(|c_{22}|^4-|c_{12}|^4)}$
is more likely to be greater than one than
$4\frac{|c_{12}|^2}{|c_{22}|^2}\frac{|c_{21}|^2}{|c_{11}|^2}$ is. This observation lets us conclude that
under secrecy constraints, TDMA is more likely to be the optimal
transmission scheme. In particular, when
\begin{align}
\left(\frac{|c_{11}|^2}{|c_{21}|^2}-\frac{|c_{21}|^2}{|c_{
11}|^2}\right)\left(\frac{|c_{22}|^2}{|c_{12}|^2}-\frac{|c_{12}|^2}{|c_{
22}|^2}\right)<4<\frac{|c_{11}|^2}{|c_{21}|^2}\frac{|c_{22}|^2}{|c_{12}|^2}
\end{align}
TDMA is preferred in secure transmissions while multiplexed
transmission is preferred when there are no secrecy limitations. In
Fig.\ref{fig:slope}, we plot the slope regions when the channel parameters are
$|c_{11}|^2=|c_{22}|^2=1,|c_{12}|^2=0.4, |c_{21}|^2=0.5$. As
explained above, secrecy slope regions are inside the slope regions
of Gaussian interference channel with no secrecy constraints. For secure transmissions, the region of TDMA is larger than that of multiplexed transmission while for
transmissions without secrecy, the region of multiplexed transmission is
larger. In Fig. \ref{fig:slope1}, we plot the slope regions when the  channel
parameters are $|c_{11}|^2=|c_{22}|^2=1,|c_{12}|^2=0.1, |c_{21}|^2=0.2$. Here, we note that
multiplexed transmission scheme is superior to TDMA scheme with and without secrecy constraints.

\section{Conclusion}\label{conclu}
In this paper, we have studied the achievable secrecy rates over Gaussian
interference channel for TDMA, multiplexed and artificial noise
schemes. Although usually TDMA  has the worst performance \cite{Liu},
we have noted that only TDMA can achieve positive secrecy rates for both users in the high-SNR
regime. In the low-power regime, we have shown that TDMA is optimal
when
$\frac{4|c_{11}|^2|c_{12}|^2|c_{22}|^2|c_{21}|^2}{(|c_{11}|^4-|c_{21}|^4)(|c_{22}|^4-|c_{12}|^4)}\geqslant
1 $. We have also shown that secrecy constraints introduce
penalty in both the minimum bit energy and slope. Finally, we have shown that TDMA is
more likely to be optimal in the presence of secrecy limitations.

\end{document}